\documentclass[preprint, showpacs, superscriptaddress, aps, pre, longbibliography]{revtex4-1}

\usepackage{amsmath}
\usepackage{graphicx}
\usepackage{color}
\usepackage{amssymb}
\usepackage{subfigure}
\usepackage{verbatim}
\usepackage{physics}
\usepackage{hyperref}
\hypersetup{
    colorlinks=true, %set true if you want colored links
    linktoc=all,     %set to all if you want both sections and subsections linked
    linkcolor=blue,  %choose some color if you want links to stand out
}

\begin{document}

\title{Scaling of causal neural avalanches in a neutral model}
\author{Sakib Matin}
\email{sakibm93@bu.edu}
\affiliation{Department of Physics, Boston University, Boston, Massachusetts 02215, USA}
\author{Thomas Tenzin}

\affiliation{Department of Physics, Boston University, Boston, Massachusetts 02215, USA}

\author{W. Klein}
\affiliation{Department of Physics, Boston University, Boston, Massachusetts 02215, USA}
\affiliation{Center for Computational Science, Boston
University, Boston, Massachusetts 02215, USA}

\date{\today}

\begin{abstract}
Neural avalanches are collective firings of neurons that exhibit emergent scale-free behavior. Understanding the nature and distribution of these avalanches is an important element in understanding how the brain functions.  We study a model of neural avalanches for which the dynamics are governed by neutral theory. The neural avalanches are defined using causal connections between the firing neurons. We analyze the scaling of causal neural avalanches as the critical point is approached from the absorbing phase. By using cluster analysis tools from percolation theory, we characterize the critical properties of the neural avalanches. We identify the tuning parameters consistent with experiments. The scaling hypothesis provides a unified explanation of the power laws which characterize the critical point. The critical exponents characterizing the avalanche distributions and divergence of the response functions are consistent with the predictions of the scaling hypothesis. We use a universal scaling function for the avalanche profile to find that the firing rates for avalanches of different durations show data collapse after appropriate rescaling. We also find data collapse for the avalanche distribution functions, which is stronger evidence of criticality than just the existence of power laws. Critical slowing-down and power law relaxation of avalanches is observed as the system is tuned to its critical point. We discuss how our results motivate future empirical studies of criticality in the brain. 
\end{abstract}
\maketitle

% ================================================================= % 
\clearpage
\section{Introduction\label{sec:intro}}
Systems with many interacting units can exhibit phenomena at macroscopic scales which cannot be elucidated from their microscopic behavior~\cite{stanleyBook, wilson1979problems}. For a system at its critical point, emergent phenomena occur at all length scales and can be understood using concepts such as scaling and universality~\cite{stanleyBook, stanley1999scaling, bigKlein}. There is a growing interest in the question of whether certain biological systems operate near a critical point~\cite{munoz2018RevMod, mora2011biological, hyman2014liquid, vicsek1995novel, lo2004common}. One question that has received much attention is the question of whether the brain operates near a critical point~\cite{munoz2018RevMod, chialvo2010emergent, Kessenich2016Synaptic, hahn2017spontaneous, dalla2019modeling, petermann2009spontaneous, priesemann2014spike, shriki2013neuronal, scarpetta2014alternation, Haimovici2013Brain, millman_brain-SOC, levina2019Critical}, commonly referred to as the criticality hypothesis~\cite{munoz2018RevMod}. Experiments have shown that neural avalanches \textit{in vivo} and \textit{in vitro} can exhibit scale-free behavior similar to thermal systems near the critical point~\cite{lo2002dynamics, beggs2003neuronal, PRX_Plasticity, friedman2012universal, Fontele_PRL19, clement2008cyclic}. The interest in the criticality hypothesis has been amplified by arguments that criticality in the brain may benefit memory storage and information processing~\cite{langton1990computation, bertschinger2004real,Stoop2016Auditory, munoz2018RevMod, honey2012slow, gautam2015maximizing, massobrio2015criticality, Haldeman2005Critical, shew2015adaptation, kinouchi2006optimal, boedecker2012information}.

The temporal-proximity binning method of defining neural avalanches~\cite{beggs2003neuronal, PRX_Plasticity, friedman2012universal, Fontele_PRL19, clement2008cyclic, ribeiro2014undersampled, levina2017subsampling} can show discrepancies from the true behavior of avalanches especially when multiple avalanches propagate through the system~\cite{MNM_OP, villegas2019timeseries}. \textit{Martinello et al.}~\cite{MNM_OP} partly addressed this issue by defining avalanches using causal-connections between firing neurons. Additionally, \textit{Martinello et al.}~\cite{MNM_OP}, studied a minimal model of neural avalanches, which is the contact process~\cite{Hinrichsen2000} where multiple neutral causally connected avalanches can propagate concurrently. We also study this neutral contact process here. The avalanche distributions in the active phase of the neutral contact process were studied in Ref.~\cite{MNM_OP}. In this paper we study the scaling of causal neural avalanches as the the critical point is approached from the absorbing phase. Our analysis reveals the relevant scaling fields or tuning parameters in the absorbing phase. We show that the causal neural avalanches in the absorbing phase are consistent with the scaling hypothesis. Additionally, we discuss how our analysis can motivate future experiments. 

The remainder of the paper is structured as follows. In Sec.~\ref{sec:model}, we outline the neutral contact process and discuss the connections to experimental studies of criticality in neural systems. We provide a brief pedagogical introduction to cluster scaling methods in Sec.~\ref{sec:theory}. In Sec.~\ref{sec:AvalancheDistribution}, we measure the critical exponents $\tau$ and $\sigma$ which characterize the scale-free causal avalanche distributions at the critical point. We approach the critical point from the absorbing phase. We also find data collapse of the distributions of avalanche size and duration near the critical point. In Sec.~\ref{sec:Response} we study how the response function diverges as a power law as the critical point is approached. We show that our measured critical exponents are consistent with the scaling hypothesis in Sec.~\ref{sec:SR}. We find that the relevant scaling field in the neutral contact process are consistent with the different tuning parameters in experiments~\cite{PRX_Plasticity, shew2009neuronal, beggs2003neuronal, poil2012critical}. In Sec.\ref{sec:SizeDur}, we analyze the scaling relation between the avalanche size and duration. In Sec.~\ref{sec:Universality}, we use scaling arguments to derive the universal avalanche profile and show data collapse for the firing rates for avalanches of different durations.  Critical slowing down is analyzed in Sec.~\ref{sec:slowDown}. In Sec.~\ref{sec:relax}, we analyze the universal relaxation dynamics in the neutral contact process near the critical point. Lastly, in Sec.~\ref{sec:end} we discuss our results, which indicate that the neutral contact process is consistent with the predictions of the scaling hypothesis. Additionally, we discuss how our results may inform future empirical studies of neural systems. 

% ================================================================= % 
\clearpage
\section{\label{sec:model}Model}
The brain is a complex system consisting of many interacting neurons. 
In the resting state, neurons have intrinsic voltages which fluctuate around some residual value. A neuron, triggered by some stimuli (endogenous or external), sends its action potential or spikes to its connected neighbors. A recipient neuron may also fire and send its spike to the connected neighbors thereby resulting in an avalanche. The firing neurons in this avalanche are causally connected. \textit{Martinello et al.}~\cite{MNM_OP},  defined neural avalanches in the neutral contact process using causal-connections and showed that this definition of neural avalanches does not suffer the ambiguities commonly found in the temporal-proximity binning method~\cite{villegas2019timeseries}. Multiple neural avalanches can propagate concurrently in the neutral contact process because of the causally connected definition. The avalanches are neutral~\cite{MNM_OP, Kimura1991, Azaele2016RevModNeutral} or symmetric because the rates that describe the dynamics of the avalanches are the same for all labels, which distinguish the different avalanches.

\begin{figure}[h]
\centering
  \includegraphics[width=0.3\linewidth]{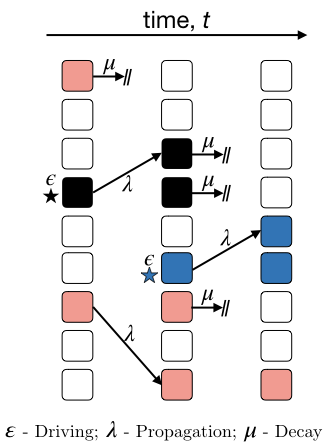}
\caption{\textbf{Neutral theory describes the dynamics of causal neural avalanches}. The boxes correspond to neurons and each column to the right is the system at a later time step. Different colors correspond to different causal avalanches. The rates for the dynamics are identical for all avalanches. A new avalanche is triggered at the driving rate $\epsilon$ and is marked with a star. An active neuron can trigger an inactive neuron anywhere in the system at the propagation rate $\lambda$. Both neurons share the same label as they are causally connected. An active neuron becomes inactive at the decay rate $\mu$. For the neural avalanche labeled by black, the size $S=2$ is the total number of activated neurons and the duration $D=2$ is the time elapsed between the first activation until all the black neurons become inactive. }
\label{fig:model}
\end{figure}

The neutral contact process consists of $N$ neurons which are fully connected. Every neuron interacts with every other neuron. A neuron is either inactive, $I$, or active $A_k$, where the index $k$ denotes the avalanche label. In Fig.~\ref{fig:model}, different colors correspond to different causal neural avalanches. The stochastic dynamics of the avalanches are described by rate equations. A new avalanche with a new label is triggered at the \textit{driving rate} $\epsilon$. An avalanche increases in size at the \textit{propagation rate} $\lambda$ as inactive neurons are triggered by active neurons. Active neurons become inactive at the \textit{decay rate} $\mu$. The rate equations describing the neutral contact process~\cite{MNM_OP} are 
\begin{align}
    I & \xrightarrow[]{\epsilon}  A_{{\rm max}[k]+1}
    \\
    I + A_k & \xrightarrow[]{\lambda}  A_{k} + A_{k}
    \\
    A_k & \xrightarrow[]{\mu}  I
\end{align}
An avalanche ends when all neurons with a given label $k$ become inactive. The size $S$ of the causal neural avalanche is the number of activations and the avalanche duration $D$ is the time between the activation of the first neuron with label $k$ to when all neurons with index $k$ become inactive, as shown in Fig.~\ref{fig:model}. 

The neutral contact process captures many of the salient biological mechanisms relevant to neural avalanches. In neural systems, the ratio of inhibitory to excitatory neurons~\cite{shew2009neuronal, beggs2003neuronal, poil2012critical} and the spontaneous triggering rate~\cite{PRX_Plasticity, priesemann2013neuronal} affect the statistics of neural avalanches. In the neutral contact process, we can tune the propagation rate, $\lambda$, and the decay rate, $\mu$, to achieve an analogous result to varying the ratio of inhibitory and excitatory neurons in experiments. In addition, the driving rate $\epsilon$ in the neutral contact process is analogous to the spontaneous triggering of neurons~\cite{PRX_Plasticity, meisel2013fading}. For these reasons, we can explore the criticality of neural avalanches in the neutral contact process in a way that is comparable to experiments.

In this paper, we only study causal neural avalanches in the absorbing phase of the neutral contact process, where $\mu\geq \lambda$ and $\epsilon\geq 0$. We simulate the model using a discrete-time asynchronous random update~\cite{Hinrichsen2000, henkel2008non}. We discuss how the code is implemented. A neuron is randomly chosen at each time step. If the neuron is inactive, then a new causal avalanche is triggered with probability $\epsilon$. If the neuron is active, then it becomes inactive with probability $\mu$ otherwise another inactive site becomes part of the same causal avalanche. All our simulations are run at $N = 10^4$, similar to Ref.~\cite{MNM_OP}. The simulations were repeated for different systems sizes up to $N=10^5$ to ensure that the measured values had converged. Additionally, we confirmed that the system is effectively ergodic using the Thrimulai-Mountain metric~\cite{Thirumalai1989Ergodic,Thirumalai1990ergodic,Thirumalai1993Activated}. We collected data on time scales much larger than the mixing time (the time needed for the system to reach effective ergodicity).

% ================================================================= %
\clearpage
\section{ Theory\label{sec:theory}}
Different physical systems exhibit universal properties near the critical point~\cite{stanley1999scaling}. We can understand the nature of a critical point by analyzing the statistical properties of fluctuations or clusters in the system~\cite{stanleyBook, stauffer2018introduction, Serino2011New}. By using real-space renormalization group techniques~\cite{coniglio1980clusters, bigKlein}, we can map domains in magnetic systems undergoing a thermodynamic phase transition to clusters in percolation models undergoing a geometric phase transition. Cluster analysis methods from percolation theory have even been used to study non-equilibrium phase transitions in integrate-and-fire systems~\cite{matin2020Effective}. Reference~\cite{stauffer1979scaling} provides a summary of scaling in percolation theory. We will use similar cluster analysis methods to study the scaling of causal avalanches in the absorbing phase of the neutral contact process. 

We can obtain thermodynamic quantities from the the avalanche number density, $n_S(S)$ which characterizes the probability of an avalanche of size $S$. Near the critical point, the number density for $S\gg1$ is
\begin{align}
    n_S \sim S^{-\tau} G\left( \frac{S}{S_c}\right).
    \label{eq:numDensity}
\end{align}
The $\tau$ exponent characterizes the power law distribution at the critical point. The characteristic size $S_c$ diverges as the system approaches the critical point, and the avalanche number density becomes a power law, $n_S \sim S^{-\tau}$. The Fisher ansatz~\cite{fisher1967theory, stauffer1975violation} assumes that the scaling function $G(x)$ is given by $G(x)=\exp(-x)$.  Equation.~\ref{eq:numDensity} and the Fisher ansatz has been applied to many systems~\cite{matin2020Effective, Serino2011New} and we will show that it is consistent with the neutral contact process. 

A system is said to be scale-free when the avalanche distribution follows a true power law, $n_S\sim s^{-\tau}$. In the scale-free case, we can plot $n_S=N_0 s^{-\tau}$ on a log-log plot to find a straight line with slope $-\tau$. Deviations from the power law occur when the system is not at the critical point. The exponential cutoff corresponds to $G(x)$ and appears as a ``knee" in the log-log plot~\cite{matin2020Effective}. Distributions with a cut-off at some characteristic scale are not scale-free. The finite lattice size in simulations also sets a cut-off. 

The scaling hypothesis was originally introduced to explain the universal behavior across disparate thermodynamic systems~\cite{stanleyBook}. According to the scaling hypothesis, the asymptotic behavior of various thermodynamic functions follow power laws characterized by critical exponents. The power laws are interconnected because they are caused by the same underlying mechanism. The predictions of the scaling hypothesis have been verified in both experiments and numerical models~\cite{stanleyBook}. Later, renormalization group methods have been used to justify the scaling hypothesis~\cite{stanley1999scaling}. 

We review the scaling of thermodynamic quantities and introduce critical exponents from statistical mechanics. Response functions quantify the change in macroscopic behavior caused by changes in intensive parameters~\cite{stanleyBook}. Our definition of the response function $\chi$ is the same as in percolation theory~\cite{stauffer1979scaling, stauffer2018introduction, sahini1994applications}, namely
\begin{align}
    \chi & = \frac{\sum_S \ S^2 \ n_S}{\sum_S \ S \ n_S}.
\end{align}
The response function $\chi$ is analogous to the magnetic susceptibility in the Ising model~\cite{stanleyBook}. The response function diverges~\cite{stanleyBook} as
\begin{align}
    \chi \sim h^{-\gamma}, \  \mathrm{as} \  h \to 0,
\end{align}
where $h$ is the difference between the tuning parameter and its critical value. Divergent response functions are hallmarks of a system near its critical point~\cite{stanleyBook}. The characteristic avalanche size $S_c$ scales~\cite{stanleyBook,stauffer2018introduction} as
\begin{align}
    S_c \sim h^{-1/\sigma}. 
\end{align}

We use scaling arguments to relate the asymptotic behavior of thermodynamic quantities near a critical point. The scaling of the characteristic avalanche size $S_c$ as a function the correlation length serves as a good pedagogical example. The correlation length $\xi$ scales as 
\begin{align}
    \xi  \sim h^{-\nu}
    \ \ \implies \xi^{-1/\nu}  \sim h.
\end{align}
The characteristic avalanche size scales as 
\begin{align}
    S_c &\sim h^{-1/\sigma} \sim \left[ \xi^{-1/\nu}\right]^{-1/\sigma} \sim \xi^{1/\sigma \nu}.
\end{align}
We will show that similar scaling arguments hold for the causal avalanches in the neutral contact process.

% ================================================================= % 
\clearpage
\section{Scale-free avalanche distribution\label{sec:AvalancheDistribution}}
In this section, we study the causal avalanche size and duration distributions in the absorbing phase of the neutral contact process where $\mu \geq \lambda$ and $\epsilon\geq 0$.  

\begin{figure}[h]
\centering 
    \includegraphics[]{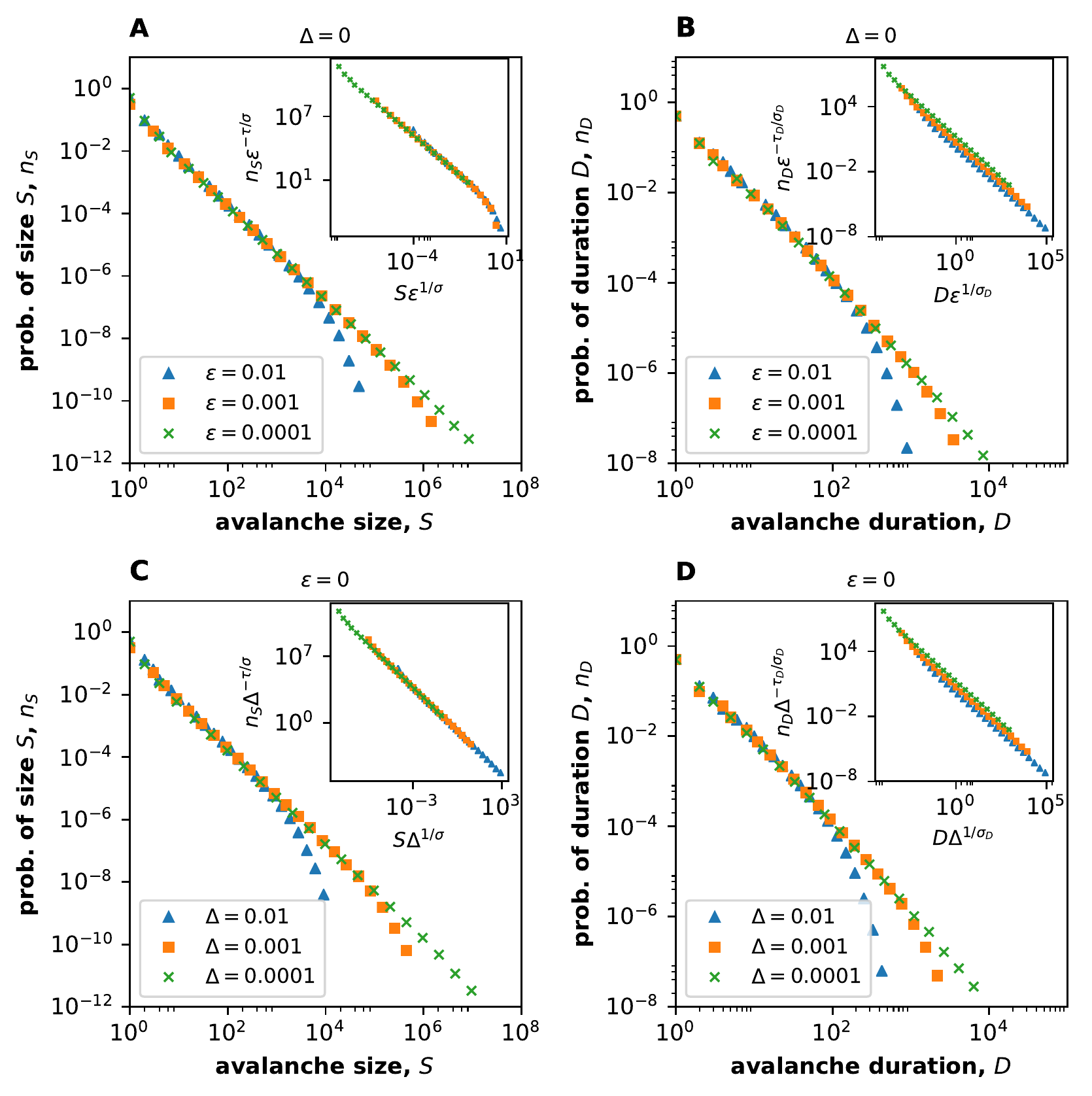}
\caption{\textbf{The distribution of avalanche sizes and durations follows a power law at the critical point, $\Delta=0$ and $\epsilon=0$.} There is exponential suppression of large avalanches for $\Delta>0$ and $\epsilon>0$. The various exponents are the same for both tuning parameters. The inset shows data-collapse for the causal avalanche distributions. In the \textbf{top row}, $\Delta=0$ and $\epsilon$ is varied. (A) The exponents for the avalanche size distribution $\tau = 1.53 \pm 0.05$. (B) The avalanche duration distribution is characterized by  $\tau_D=1.92\pm0.11$. In the \textbf{bottom row}, $\Delta$ is varied for $\epsilon=0$. (C) The corresponding exponent $\tau = 1.51 \pm 0.05$. (D)$\tau_D = 1.93 \pm 0.11$ is the critical exponent for the avalanche duration distribution.  }
\label{fig:scaling}
\end{figure}

One of the challenges of the criticality hypothesis is that the tuning parameters in real neural systems are not known. Different experiments have suggested different tuning parameters~\cite{PRX_Plasticity, shew2009neuronal, poil2012critical}. In Ref.~\cite{PRX_Plasticity}, the experiments \textit{in vitro} and \textit{in vivo} show that the spontaneous triggering rate of neurons may be interpreted as a tuning parameter. This parameter corresponds to $\epsilon$ in the neutral contact process. The experiments reported in Ref.~\cite{shew2009neuronal, beggs2003neuronal, poil2012critical} use pharmacological means to alter the excitation-inhibition ratio to alter the proximity to the critical point. We can achieve similar results by varying the rates $\lambda$ and $\mu$. Our analysis shows that the relevant scaling fields for the neutral contact process depends on $\epsilon$ and propensity $\Delta = \mu - \lambda$, which are consistent with the different experimental results~\cite{PRX_Plasticity,shew2009neuronal, beggs2003neuronal, poil2012critical}. We find that the critical point for the neutral contact process is $\Delta =\mu - \lambda=0$ and $\epsilon=0$. We find  that the distributions of avalanche sizes and durations follows power laws for $\Delta=0$ and $\epsilon=0$. Larger avalanches are suppressed for $\epsilon>0$ and $\Delta>0$. 

\begin{figure}[h]
\centering 
    \includegraphics[]{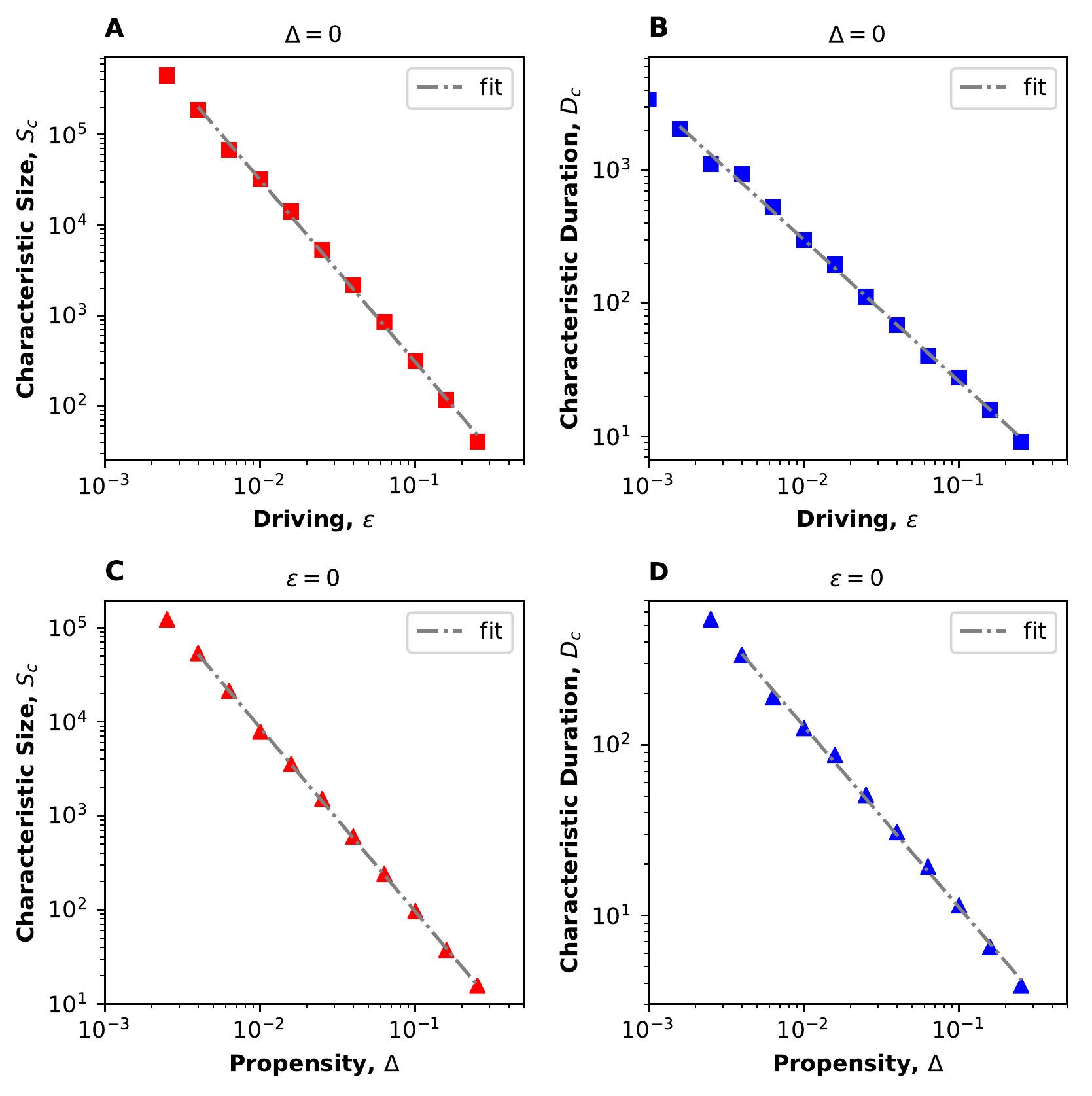}
\caption{ \textbf{The characteristic avalanche size $S_c$ and duration $D_c$ diverge as the critical point is approached.} For each value of $\Delta$ or $\epsilon$, we fit the avalanche size and duration distributions to Eq.~\ref{eq:sizeScaling} and Eq.~\ref{eq:durationScaling} to obtain $S_c$ and $D_c$ respectively. In the \textbf{top row}, $\epsilon$ is varied at fixed $\Delta=0$. (A) The characteristic avalanche size $S_c$ diverges with critical exponent $\sigma=0.54 \pm 0.07 $. (B) The characteristic duration $D_c$ diverges with critical exponent $\sigma_D = 0.95 \pm 0.06$. In the \textbf{bottom row}, $\Delta$ is varied at fixed $\epsilon=0$. (C) The measured exponent for $S_c$ is $\sigma = 0.52 \pm0.08$. (D) The characteristic duration diverges with the exponent $ \sigma_D = 1.02 \pm 0.09$.}
\label{fig:characteristic}
\end{figure}

In Fig.~\ref{fig:scaling} we observe that the distribution of the avalanche sizes $n_S$ and duration $n_D$ satisfy power law at the critical point with the exponents $\tau$ and $\tau_D$. As we tune the system away from the critical point by increasing the driving $\epsilon$ or the propensity $\Delta$, we find  exponential suppression of the large avalanches characterized by the exponents $\sigma$ and $\sigma_D$ for the size and duration respectively. The distribution functions for the avalanche size and duration are
\begin{align}
    n_S & \sim S^{-\tau}\exp[-\frac{S}{S_c}], 
    \label{eq:sizeScaling}
    \\
    n_D & \sim D^{-\tau_D}\exp[-\frac{D}{D_c}], 
    \label{eq:durationScaling}
\end{align}
where, $S_c$ is the characteristic avalanche size and $D_c$ is the characteristic duration size~\cite{fisher1967theory, stauffer1979scaling, stauffer2018introduction}. For critical propensity $\Delta = 0$, the characteristic avalanche size scales as $S_c \sim \epsilon^{-1/\sigma}$ and the characteristic duration scales as $D_c \sim \epsilon^{-1/\sigma_D} $. When $\epsilon=0$ the scaling is $S_c \sim  \Delta^{-1/\sigma}$ and $D_c \sim \Delta^{-1/\sigma_D}$. 

Our measured value of $\tau$ and $\tau_D$ in Fig.~\ref{fig:scaling} are consistent with the theoretical mean field values, $\tau^{\rm MF}=1.5$ and $\tau_D^{\rm MF}=2$~\cite{munoz2018RevMod} and experimentally reported values~\cite{PRX_Plasticity}. In Fig.~\ref{fig:characteristic} we fit the exponents, $\sigma$ and $\sigma_D$ which characterize the exponential suppression of large avalanches. The critical exponents $\sigma$ and $\sigma_D$ for neural systems have not been reported in experiments.

A remarkable consequence of the scaling hypothesis is the existence of universal scaling functions which are usually obtained via data collapse~\cite{stanleyBook, stanley1999scaling, sethna2001crackling}, where results for different values of the control parameters collapse on to a single curve after appropriate rescaling. In Fig.~\ref{fig:scaling}, the insets show data collapse for the distributions of the causal avalanches. In Fig.~\ref{fig:scaling}A, the causal avalanche size distribution for different values of $\epsilon$ is plotted. The scaling form is 
\begin{align}
    n_S &\sim S^{-\tau}\exp\left(-\frac{S}{S_c} \right) \sim S^{-\tau}\exp\left(-S \epsilon^{1/\sigma} \right).
\end{align}
We multiply both sides by $\epsilon^{-\tau/\sigma}$, 
\begin{align}
    n_S \epsilon^{-\tau/\sigma} & \sim  \epsilon^{-\tau/\sigma} S^{-\tau}\exp\left(-S \epsilon^{1/\sigma} \right)
    \\
    n_S \epsilon^{-\tau/\sigma} &\sim  \left( S \epsilon^{1/\sigma}   \right)^{-\tau}\exp\left(-S \epsilon^{1/\sigma} \right).
\end{align}
In the inset of Fig.~\ref{fig:scaling}A, we plot $n_S \epsilon^{-\tau/\sigma} $ as a function of the rescaled avalanche size $S \epsilon^{1/\sigma}$ to find data-collapse. The exact same scaling arguments be used to derive the rescaled scaled variables in Fig.~\ref{fig:scaling}B, \ref{fig:scaling}C and \ref{fig:scaling}D, which are as follows
\begin{align}
    n_D \epsilon^{-\tau_D/\sigma_D} &\sim  \left( D \epsilon^{1/\sigma_D}   \right)^{-\tau_D}\exp\left(-D \epsilon^{1/\sigma_D} \right)
    \\
    n_S \Delta^{-\tau/\sigma} &\sim  \left( S \Delta^{1/\sigma}   \right)^{-\tau}\exp\left(-S \Delta^{1/\sigma} \right)
    \\
    n_D \Delta^{-\tau_D/\sigma_D} &\sim  \left( D \Delta^{1/\sigma_D}   \right)^{-\tau_D}\exp\left(-D \Delta^{1/\sigma_D} \right).
\end{align}
The data collapse of the causal avalanche distributions in the insets of Fig.~\ref{fig:scaling} is compelling evidence relevant scaling fields or tuning parameters in the absorbing phase depends on $\epsilon$ and $\Delta$.

% ================================================================= %
\clearpage
\section{\label{sec:Response} Divergent Response Function}
\begin{figure}[h]
\centering 
    \includegraphics{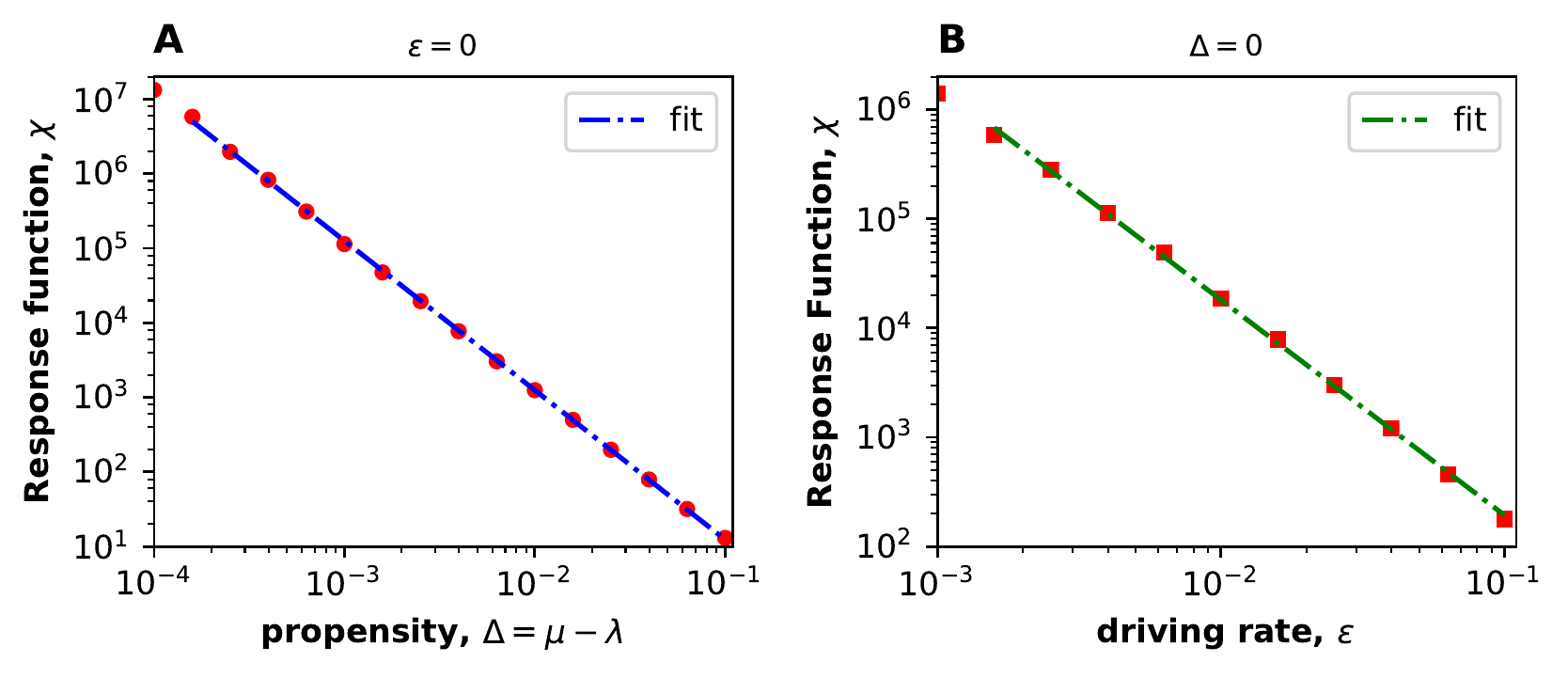}
\caption{\textbf{The response function $\chi$ diverges as the system approaches the critical point.} The exponent $\gamma$ characterizes the divergence and is the same when the critical point is approached at constant driving $\epsilon=0$ or $\Delta=0$. (A) For $\epsilon=0$, when $\Delta$ is varied the exponent is $\gamma=2.00 \pm 0.02$ (B) Varying the driving rate $\epsilon$, the exponent is $\gamma=1.97\pm0.04$ for $\Delta=0$. }
\label{fig:response}
\end{figure}

Divergent response functions are hallmarks of critical points~\cite{stanleyBook}. We consider response function $\chi$, which is from percolation theory, is given by~\cite{stauffer1979scaling, matin2020Effective}
\begin{align}
    \chi & = \frac{\sum_S S^2 \  n_S}{\sum_S S \  n_S}.
    \label{eq:response}
\end{align}
In Fig.~\ref{fig:response}, the exponent $\gamma$ is found to be the same whether the critical point is approached by decreasing $\Delta$ at $\epsilon=0$ or by decreasing $\epsilon$ at $\Delta=0$. When we vary $\Delta$ we measured $\gamma = 2.00\pm0.02$ and when we vary $\epsilon$ the exponent is $\gamma=1.97\pm 0.04$.

Our results indicate that the response function in the neutral contact process diverges as we vary $\epsilon$ or $\Delta$ in the absorbing phase.

% ================================================================= %
\clearpage
\section{Scaling Relations\label{sec:SR}}
The scale-free behavior at critical points can be attributed to underlying singularities in theromodynamic functions~\cite{fisher1967theory, stanleyBook}. We use scaling theory to relate the different critical exponents in the neutral contact process.

According to the scaling hypothesis, the response functions can be described by generalized homogeneous functions near the critical point~\cite{stanleyBook}. We can write $\chi$ in Eq.~\ref{eq:response} as
\begin{align}
    \chi  & = \frac{\sum_S S^2 n_S}{\sum_S S n_S},
    \\
    & = \frac{\int_1^\infty \ S^{2-\tau} \ G(S/S_c) \ \mathrm{d} S }{\int_1^\infty \ S^{1-\tau} \ G(S/S_c) \ \mathrm{d} S}.
\end{align}
We make change of variables $u=S/S_c$ to find
\begin{align}
    \chi  & = \frac{S_c^{3-\tau} \int_{1/S_c}^{\infty} \ u^{2-\tau} \ G(u) \ \mathrm{d}u  }{S_c^{2-\tau} \int_{1/S_c}^{\infty} \ u^{1-\tau} \ G(u) \ \mathrm{d}u },
\end{align}
and set $G(u)=\exp(-u)$~\cite{fisher1967theory} to express $\chi$ as
\begin{align}
    \chi & = S_c \frac{\Gamma(3-\tau)}{\Gamma(2-\tau)},
\end{align}
where $\Gamma$ is the gamma function. 

Near the critical point, we find that the response function scales as $\chi \sim S_c$, where the characteristic size $S_c$ scales as $\epsilon^{-1/\sigma}$. The scaling relation between the critical exponents are
\begin{align}
      \epsilon^{-1/\sigma} \sim \epsilon^{-\gamma} \implies \gamma=1/\sigma.
    \label{eq:responseScaling}
\end{align}
Our measured exponents in Figs.~\ref{fig:characteristic} and~\ref{fig:response} are consistent with Eq.~(\ref{eq:responseScaling}). The results are the same when we vary $\Delta$ at $\epsilon=0$. Our derivation is similar to scaling arguments used in percolation theory~\cite{stauffer1979scaling}, except that the denominator of the response function also contributes to the divergence. 

% ================================================================= %
\clearpage
\section{Size-Duration scaling of avalanches\label{sec:SizeDur}}
We can relate the size of the causal avalanches to their duration using  scaling arguments similar to Ref.~\cite{sethna2001crackling}. The critical point is approached by varying $\epsilon$ for $\Delta=0$. The average avalanche size follows the same scaling as the characteristic avalanche size, $\langle S(D) \rangle \sim \epsilon^{-1/\sigma}$. The correlation length scales as $\xi\sim\epsilon^{-\nu}$. The dynamic critical exponent $z$ relates the duration $D$ to $\xi$~\cite{CriticalExpDP, Roli2018Dynamical}. The scaling of D is
\begin{align}
    D & \sim \xi^{z} \sim \epsilon^{-\nu z},
    \\
    D^{-1/\nu z} & \sim \epsilon.
\end{align}
Now we can relate the average size $\langle S(D)\rangle$ to the duration as 
\begin{align}
    \langle S(D) \rangle & \sim \epsilon^{-1/\sigma} \sim 
    \left[ D^{-1/\nu z}\right]^{-1/\sigma},
    \\ 
    & \sim D^{1 /(\sigma \nu z)}.
\end{align}
The same scaling arguments can be used when $\Delta$ is varied for $\epsilon=0$. 

\begin{figure}[h] 
\centering 
    \includegraphics{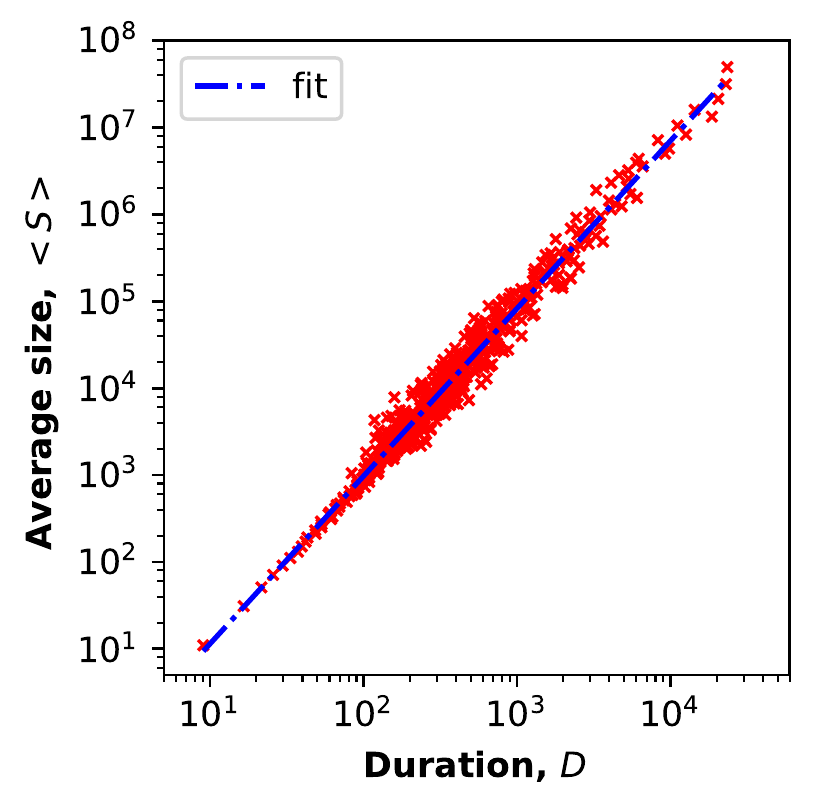}
\caption{\textbf{The scaling of the average avalanche size as a function of the duration at the critical point is consistent with the scaling laws.} At the critical point the scaling is $\langle S(D) \rangle\sim D^{1/\sigma \nu z}$. The numerical estimate of $\frac{1}{\sigma \nu z} = 1.96 \pm 0.03$ is consistent with Eq.~\ref{eq:sizeDuration}. }
\label{fig:sizeDur}
\end{figure} 

The critical exponents for the distribution of avalanche size and duration can be related to the exponent $\sigma \nu z$, by the identity~\cite{sethna2001crackling} 
\begin{align}
    \frac{\tau_D-1}{\tau-1}=\frac{1}{\sigma\nu z }.
    \label{eq:sizeDuration}
\end{align}
We measure the exponents on the left and right side of Eq.~\ref{eq:sizeDuration} independently. From Fig.~\ref{fig:sizeDur} the measured value of $1/(\sigma \nu z) = 1.93 \pm 0.03$ is consistent with the measured values of $(\tau_D-1)/(\tau-1)$ in Fig.~\ref{fig:scaling}.  The scaling relation between the size and duration of neural avalanches have been verified experimentally~\cite{Fontele_PRL19, friedman2012universal}. The predictions of the scaling hypothesis provide stricter criteria for criticality than just the existence of power law distributions.

% ================================================================= % 
\clearpage
\section{Universal Avalanche Profile \label{sec:Universality}}
\begin{figure}[h]
\centering 
    \includegraphics[]{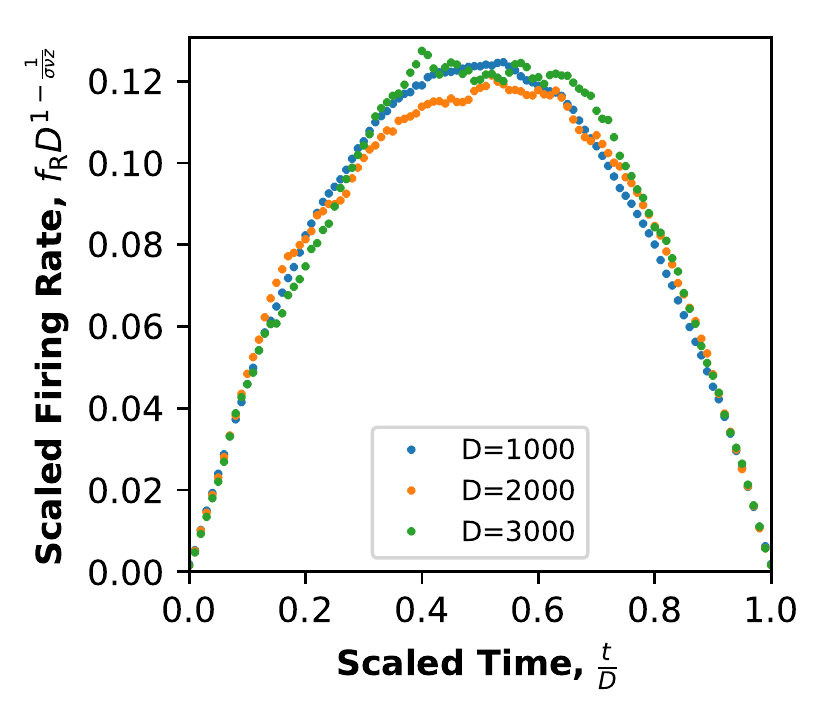}
\caption{ \textbf{Neural avalanches have a universal avalanche profile at the critical point.} The firing rate scaled by $D^{1-1/(\sigma \nu z)}$ as a function of the rescaled time $t/D$ shows data collapse for avalanches of different durations.}
\label{fig:profile}
\end{figure}

The avalanche profile, which describes the firing rate as a function of time, can be described by a universal scaling function near the critical point. The firing rate corresponds to the number of activations per unit time. From the scaling hypothesis, we assume the average firing rate is described by a generalized homogeneous function, which can be written as $f_{\rm R}(t,D) = D^b f_{\rm R}(t/D)$~\cite{stanleyBook, sethna2001crackling, gleeson2017temporal, ponce2018whole}. We compute exponent the exponent $b$, 
\begin{align}
    \langle S(D)  \rangle  & = \int  f_{\rm R} (t,D) dt  = \int D^{b}f_{\rm R} (t/D) dt \sim D^{b+1}.
\end{align}
By using the scaling identity $\bm{\langle} S(D) \bm{\rangle} \sim D^{1/(\sigma \nu z)}$, we find $b=1/(\sigma \nu z) - 1$. Figure~\ref{fig:profile} shows the data collapse for avalanches of different durations, when we scale the firing rate by $D^{1-1/(\sigma \nu z)}$ and plot it as a function of the rescaled time $t/D$. Our derivation follows Ref.~\cite{sethna2001crackling}. 

\clearpage

Data collapse is an impressive example of universality in neural avalanches. Universal scaling functions can be used as strict criteria for criticality because the data collapse is observed only sufficiently close to the critical point. Under certain circumstances, data collapse for the avalanche profile has been reported for \textit{in vitro} experiments~\cite{friedman2012universal, yu2013universal,Fontele_PRL19}, where the avalanches are defined using the temporal proximity binning method. Our result shows that causal avalanches in the absorbing phase of the neutral contact process follow similar scaling behavior.

% ================================================================= % 
\clearpage
\section{Critical Slowing down\label{sec:slowDown}}
A well-known consequence of criticality is a time scale that diverges as the critical point is approached~\cite{Hohenberg1977RevMod}. This is known as critical slowing down. Here, we analyze how the time to reach a stationary state diverges as the system is tuned to the critical point.

\begin{figure}[h]
\centering 
    \includegraphics[]{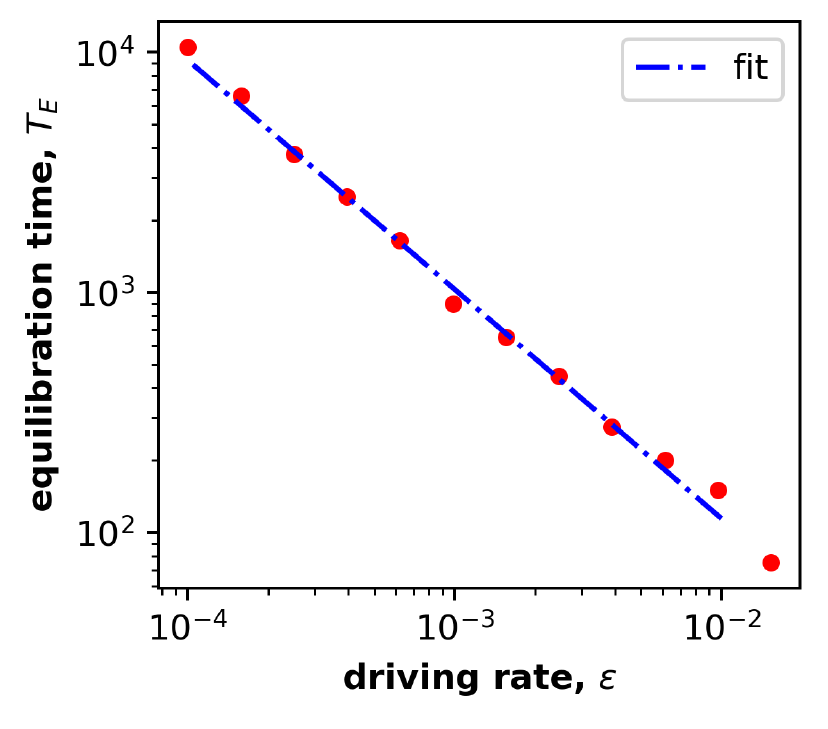}
\caption{\textbf{The equilibration time $T_{\rm E}$ diverges as the system approaches the critical point $\epsilon \to 0$ with $\Delta = 0$.} $T_{\rm E}$ is the the time for $U(t)$ to reach a steady-state value when the neutral contact process is initiated with a single active neuron. We find $T_{\rm E} \sim \epsilon^{-\nu z}$. The measured value, $\nu z=0.95\pm0.04$, is consistent with scaling arguments.}
\label{fig:equilibration}
\end{figure}

We study the equilibration time for neutral contact process after initializing with a single active neuron. We analyze the dynamics of $U(t)$, which is the number of unique causal avalanches at time $t$. $U(t)$ is a population level quantity because it is computed using information from the whole system at a particular instance in time. We determine that the system has reached a stationary state when $U(t)$ reaches a constant rolling time average. In Fig.~\ref{fig:equilibration} we plot the time for $U(t)$ to reach a steady state $T_{\rm{E}}$ as a function of the driving rate $\epsilon$ at $\Delta=0$. We can use scaling arguments to relate the dynamic exponent $z$ to the other critical exponents; the correlation length and time scale as $\xi \sim \epsilon^{-\nu}$ and $t \sim \xi^z$ respectively. In Fig.~\ref{fig:equilibration} the system approaches the critical point, $T_{\rm E}$ diverges as
\begin{align}
    T_{\rm E} \sim \xi^{z} \sim \epsilon^{-\nu z}.
\end{align} 

Our measured dynamic critical exponent $\nu z=0.95\pm0.04$ is consistent with Eq.~\ref{eq:sizeDuration} and $1/(\sigma \nu z)$, where $\sigma$ is determined in Fig.~\ref{fig:characteristic}. The measured exponent remains the same when we repeat the analysis using the total activity, $\rho$, instead of $U$. Hence, the neutral contact process exhibits a divergent time scale characteristic of critical systems and is consistent with the scaling hypothesis.

% ================================================================= % 
\clearpage
\section{Relaxation Dynamics \label{sec:relax}}
\begin{figure}[h!]
\centering 
    \includegraphics[width=1\linewidth]{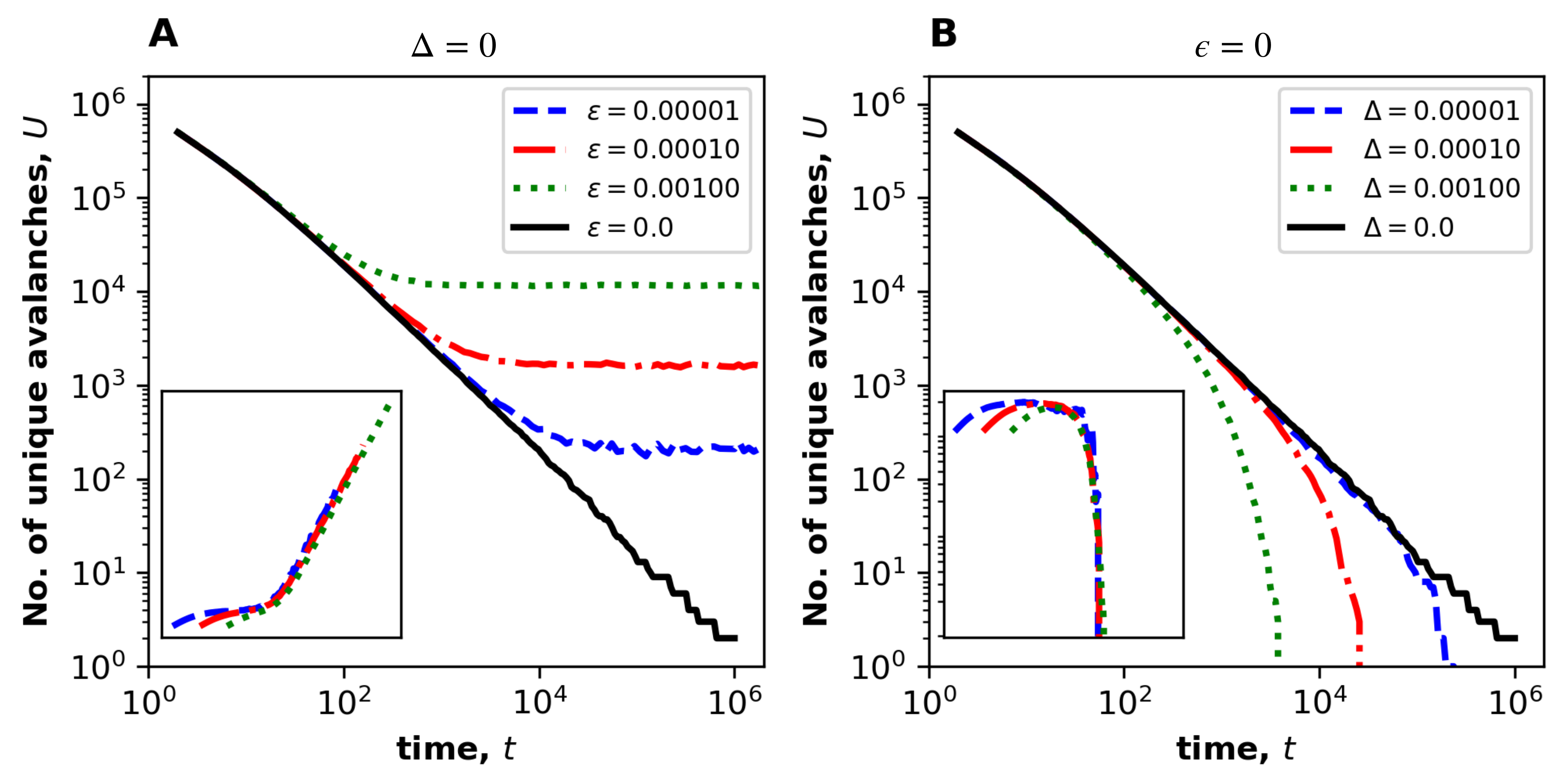}
\caption{\textbf{The number of unique causal neural avalanches $U(t)$ decays as a power law, $U(t)\sim t^{-\alpha}$, at the critical point.} The measured critical exponent $\alpha=0.99\pm 0.04$ matches mean-field value $\alpha_{\rm MF}=1$. (A) For $\epsilon>0$ and critical propensity $\Delta=0$, $U(t)$ reaches a fluctuating state. (B) For sub-critical propensity $\Delta>0$ and $\epsilon=0$, $U(t)$  decays exponentially to the absorbing state. Inset plots show data collapse for the rescaled variables. }
\label{fig:relax}
\end{figure}

Power law temporal relaxation is a hallmark of critical systems~\cite{Hohenberg1977RevMod}. We initialize the neutral contact process with every neuron active and belonging to a unique causal avalanche and analyze the relaxation to either a fluctuating state or to an absorbing (inactive) state. In Fig.~\ref{fig:relax}A, we vary $\epsilon$ for $\Delta=0$ and analyze how the system decays to a fluctuating state. In Fig.~\ref{fig:relax}B the system decays to an absorbing state as we have set the driving rate $\epsilon=0$. We find that the number of unique avalanches, $U(t)$, decays as a power law for the critical value $\Delta=0$, and exponentially for $\Delta>0$. The critical exponent $\alpha$ characterizes the power law relaxation. Our measured value is $\alpha=0.99\pm 0.04$ in Fig.~\ref{fig:relax} and is consistent with the mean-field value $\alpha_{\rm MF}=1$~\cite{Hinrichsen2000}. Inset plots in Fig.~\ref{fig:relax} show data collapse for the relaxation of $U(t)$ by plotting the rescaled variables $U \rightarrow U t^\alpha$ as a function of  $t \to (t|\epsilon|)^{\nu z}$ and $t \to (t|\Delta|)^{\nu z}$ respectively in Figs.~\ref{fig:relax}A and B. This technique has been used in the study of directed percolation, which exhibits a nonequilibrium phase transition~\cite{henkel2008non, Hinrichsen2000}. The data was averaged over $10^6$ iterations. 

Our analysis of other dynamical properties in the neutral contact process also support criticality. For finite values of the driving rate $\epsilon$ the system evolves to a stationary state. The time-averaged value of the number of unique clusters scales as $\langle U \rangle \sim \epsilon^\lambda$ where $\lambda = 0.80 \pm 0.02 $. The divergence in the time scale to reach the stationary state was analyzed in Sec~\ref{sec:slowDown}.

By studying the dynamics of population level quantities, we have found characteristic power laws at the critical point. Furthermore, the data collapse for the decay of $U(t)$ highlights the universal dynamics in the neutral contact process near the critical point.

% ================================================================= % 
\clearpage
\section{Discussion\label{sec:end}}
Scale-free neural avalanches \textit{in vivo} and \textit{in vitro} are a remarkable emergent phenomena which have intrigued physicists~\cite{munoz2018RevMod} and neuroscientists~\cite{beggs2012being, clawson2017adaptation, shew2013functional}. The theory of critical phenomena is a promising explanation of the scale-free behavior~\cite{friedman2012universal, Fontele_PRL19, Levina2009Phase}, and is further motivated by the arguments that criticality in the brain may have functional advantages~\cite{shew2013functional}. 

We have analyzed the scaling of causal neural avalanches in the absorbing phase of the neutral contact process. As we allow the different parameters to approach their critical values, the response function exhibit a power law divergence, similar to thermal systems~\cite{stanleyBook}. Additionally, the causal neural avalanches show scale-free distributions for $\Delta=0$ and $\epsilon=0$. Large causal neural avalanches are exponentially suppressed when $\epsilon>0$ and $\Delta>0$. The values of the critical exponents $\tau$ and $\tau_D$ in Ref.~\cite{MNM_OP} are consistent with our measured values. We used cluster analysis techniques to study the causal avalanches and measure critical exponents, $\sigma$, $\sigma_D$, $\gamma$ and $\nu z$.
Our measured exponents obey scaling laws in Eq.~\ref{eq:responseScaling}. We use scaling arguments to relate the average avalanche size to the duration and numerically verify the result. 

We construct a strict criteria for criticality in the neutral contact process using the scaling hypothesis. A striking prediction of the scaling hypothesis is the existence of universal scaling functions. In the neutral contact process the avalanche profile shows data collapse after appropriate rescaling and has been observed in experiments~\cite{friedman2012universal}. We also find data collapse for the avalanche size and duration distribution. We showed that the dynamics of population level observables in the neutral contact process are also consistent with criticality. When we initialize the system with a single active neuron, the steady state is reached over some characteristic time. As the critical point is approached, we find there is a divergent time scale. The dynamic critical exponent is consistent with scaling arguments. We analyze how the system relaxes from a maximally diverse state. We find deviations from power law decay depend on the distance from the critical point. By using scaling arguments, we find data collapse for the relaxation dynamics.

Our analysis of the neutral contact process shows that the relevant scaling field in in the absorbing phase depends on $\epsilon$ and $\Delta$. The results of Ref.~\cite{MNM_OP} suggest that relevant scaling field depends only on $\epsilon$ in the active phase of the neutral contact process, where $\Delta<0$. An important question is about the correspondence between the tuning parameters in experiments and those in the neutral contact process. The experiments in Ref.~\cite{PRX_Plasticity} imply that the control parameter may be the spontaneous triggering rate, which corresponds to the driving rate $\epsilon$ in the neutral contact process. In separate experiments~\cite{shew2009neuronal, beggs2003neuronal, poil2012critical}, the excitation-inhibition ratio was varied to tune the system toward criticality, which we achieved by varying the propensity, $\Delta$ in the neutral contact process. However, in experiments the neural avalanches are defined using temporal proximity binning method which can be different from the causal avalanche definition used by us and Ref.~\cite{MNM_OP}. Our results emphasize the need to incorporate causal connections in future experiments studying neural avalanches as pointed out by Ref.~\cite{MNM_OP}. Inferring causal information in real neural systems remains an open problem. References~\cite{williams2017unveiling, stavroglou2020unveiling} provide promising steps to to address this issue. Additionally, Ref.~\cite{Zierenberg2020Description} has studied how coalescence in the microscopic dynamics can affect the macroscopic observeables in the branching process.

Our results may motivate future experimental studies of neural avalanches. The divergence of the response function can be used to identify the critical point in real neural systems. We outline a possible way to experimentally verify the scaling relation $\gamma=1/\sigma$. In an experiment, neural avalanches are recorded for different values of the tuning parameter $f$. The avalanche distribution function can be fit to $n_S \sim S^{-\tau} \exp\left[-S/S_c(f) \right]$, where $S_c(f)$ is the characteristic avalanche size for a given value of $f$. Additionally, the response function $\chi(f)$ can also be computed. We rescale the tuning parameter $f \to \overline{f}$ such that the log-log scaled  plot of $S_c$ against $\overline{f}$ gives a straight line. Similarly, we can plot the response function $\chi$ against $\overline{f}$ on a log-log scaled plot. The measured slopes for the two graphs will be the same if the $\gamma=1/\sigma$ scaling law is obeyed. We would find better data collapse for firing rates from data sets where $S_c$ and $\chi$ are large, as the data collapse indicates that the experimental system is near the critical point. Some of predictions of our scaling may even be verified using existing neural data as the effects of coalescence is expected to be small near the critical point. Experiments in Ref.~\cite{friedman2012universal} reported data collapse for the avalanche profile in certain samples.

Numerous studies~\cite{langton1990computation, beggs2003neuronal, bertschinger2004real, munoz2018RevMod, kinouchi2006optimal} have discussed the possible functional benefits of criticality in the brain. Ref.~\cite{beggs2003neuronal} showed that information transmission is maximized for critical neural avalanches. Our results raise the important question of if there are similar benefits associated with causal neural avalanches near the critical point. Spike-timing-dependent plasticity(STDP) is a biological learning mechanism which uses causal information about firing neurons~\cite{MNM_OP,stepp2015synaptic} to update synaptic weights between neurons. STDP has been shown to be responsible for maintaining stable retrievable firing patterns~\cite{stepp2015synaptic}. We have shown that the causal neural avalanches of all scales occur at the critical point. We plan to address the question of whether criticality enhances the number or stability of STDP firing patterns by studying causal neural avalanches in integrate and fire models of neurons. 

Our work sheds new light on the scaling of causal neural avalanches. Our results indicate that the relevant scaling field in the absorbing phase of the NMN is consistent with  experiments~\cite{PRX_Plasticity, beggs2003neuronal}. Our results motivate questions for future studies and provide a promising path to a unified theory of neural avalanches. 

\begin{acknowledgements}
We would like to thank Ashish B. George and Harvey Gould for insightful comments and critical reading of the manuscript. T.T. would like to acknowledge financial support from Boston University’s Undergraduate Research Opportunities Program.
\end{acknowledgements}

\clearpage
\nocite{*}
\bibliography{References}

\end{document}